%% file: Template.tex
\documentclass{article}
\usepackage[preprint]{spconfa4}
\usepackage{cite}
\usepackage{amsmath,amssymb,amsfonts}
\usepackage{textcomp}
\usepackage[table,xcdraw]{xcolor}
\usepackage{multirow}
\usepackage{acronym} 
\usepackage{svg}
\usepackage[hypcap=true]{subcaption, caption}
\usepackage{graphicx}
\usepackage{tabularx}

\usepackage[citecolor=black,%black!40!red,
            colorlinks=true,
            linkcolor=black,%blue!10!black,
            urlcolor  = black,
            pdfpagelabels={true},
            pdftitle={Utterance Weighted Multi-Dilation Temporal Convolutional Networks for Monaural Speech Dereverberation},
            pdfauthor={William Ravenscroft, Stefan Goetze, Thomas Hain},
            pdfkeywords={speech dereverberation, temporal convolutional network, speech enhancement, receptive field, deep neural network},
            pdfsubject={speech dereverberation, temporal convolutional network, speech enhancement, receptive field, deep neural network}]
            {hyperref}
\usepackage{orcidlink}

\acrodef{AC}    {acoustic conditions}
\acrodef{ASR}   {automatic speech recognition}
\acrodef{BLSTM} {bidirectional long short term memory}
\acrodef{CB}    {convolutional block}
\acrodef{Conv-TasNet} {convolutional \ac{TasNet}}
\acrodef{CSM}   {clean speech mixtures}
\acrodef{cLN}   {channelwise layer normalization}
\acrodef{DAE}   {denoising autoencoder}
\acrodef{WD-Conv} {weighted multi-dilation depthwise-separable convolution}
\acrodef{DL}    {deep learning}
\acrodef{DNN}   {deep neural network}
\acrodef{DPRNN} {dual path recurrent neural network model}
\acrodef{DPTNet}{dual path Transformer network}
\acrodef{WD-TCN}  {weighted multi-dilation temporal convolutional network}
\acrodef{DS-Conv}{depthwise-separable convolution}
\acrodef{D-Conv}{depthwise convolution}
\acrodef{E2E}   {end-to-end}
\acrodef{ER}    {early reflection}
\acrodef{gLN}   {global layer normalization}
\acrodef{LSTM}  {long short term memory}
\acrodef{LR}    {late reflection}
\acrodef{MHA}   {multihead attention}
\acrodef{LMHSA} {linear multihead self attention}
\acrodef{MOS}   {Mean Opinion Score}
\acrodef{MR}    {mask refinement}
\acrodef{NMF}   {non-negative matrix factorization}
\acrodef{NSM}   {noisy speech mixture}
\acrodef{NRSM}  {noisy reverberant speech mixture}
\acrodef{PESQ}  {perceptual evaluation of speech quality}
\acrodef{PIT}   {permutation invariant training}
\acrodef{PM}    {post-masking}
\acrodef{PReLU} {parametric rectified linear unit}
\acrodef{P-Conv}{pointwise convolution}
\acrodef{ReLU}  {rectified linear unit}
\acrodef{RF}    {receptive field}
\acrodef{RIR}   {room impulse response}
\acrodef{RSM}   {reverberant speech mixture}
\acrodef{SE}    {squeeze-and-excite}
\acrodef{SISDR} {scale-invariant signal-to-distortion ratio}
\acrodef{SISDRi} {\ac{SISDR} improvement}
\acrodef{SDR}   {signal-to-distortion ratio}
\acrodef{SNR}   {signal-to-noise ratio}
\acrodef{SRMR}  {speech-to-reverberation modulation energy ratio}
\acrodef{SP}    {signal processing}
\acrodef{STFT}  {short-time Fourier transform}
\acrodef{STOI}  {short-time objective intelligibility}
\acrodef{ESTOI} {extended short-time objective intelligibility}
\acrodef{TasNet}{Time-domain audio separation network}
\acrodef{TCN}   {temporal convolutional network}
\acrodef{UPGMA} {unweighted pair group method with arithmetic mean}
\acrodef{WER}   {word error rate}
\acrodef{WPE}   {weighted prediction error}

\newcommand{\mat}[1]{\mathbf{#1}}
\newcommand{\vek}[1]{\ensuremath{\mathbf{#1}}}    % (fette) vector matrices ABC...

\newcommand{\Real}{\mathbb{R}}

\copyrightnotice{978-1-6654-6867-1/22/\$31.00~\copyright2022 IEEE}
                   
\title{Utterance Weighted Multi-Dilation Temporal Convolutional Networks for Monaural Speech Dereverberation}
\name{
    {
        William Ravenscroft$^{\orcidlink{0000-0002-0780-3303}}$, Stefan Goetze$^{\orcidlink{0000-0003-1044-7343}}$ and Thomas Hain$^{\orcidlink{0000-0003-0939-3464}}$
    }
    \thanks{
        This work was supported by the Centre for Doctoral Training in Speech and Language Technologies (SLT) and their Applications funded by UK Research and Innovation [grant number EP/S023062/1]. This work was also funded in part by 3M Health Information Systems, Inc.
    }
}
\address{
    \textit{Department of Computer Science}, 
    \textit{{The} University of Sheffield}, Sheffield, United Kingdom \\
    \{jwravenscroft1, s.goetze, t.hain\}@sheffield.ac.uk \vspace*{-0.2cm}
}

\begin{document}
\ninept
\maketitle

\begin{abstract}

Speech dereverberation is an important stage in many speech technology applications. Recent work in this area has been dominated by deep neural network models. \Acp{TCN} are 
%one such 
deep learning models that have been proposed for sequence modelling in the task of dereverberating speech. 
%One property of the \ac{TCN} is that it has a fixed receptive field. 
In this work a weighted multi-dilation depthwise-separable convolution is proposed to replace standard depthwise-separable convolutions in \ac{TCN} models. This proposed convolution enables the \ac{TCN} to 
dynamically focus on more or less local information in its receptive field at each convolutional block in the network. It is shown that this \ac{WD-TCN} consistently outperforms the \ac{TCN} across various model configurations and using the \ac{WD-TCN} model is a more parameter-efficient method to improve the performance of the model than increasing the number of convolutional blocks. The best performance improvement over the baseline \ac{TCN} is 0.55~dB~\ac{SISDR} and the best performing \ac{WD-TCN} model attains 12.26~dB SISDR on the WHAMR dataset.

\end{abstract}
\begin{keywords}
speech dereverberation, temporal convolutional network, speech enhancement, receptive field, deep neural network
\end{keywords}
%
%%%%%%%%%%%%%%%%% SECTION %%%%%%%%%%%%%%%%% 
\section{Introduction}
\label{sec:intro}
%% General introduction
Speech dereverberation remains an important task for robust speech processing \cite{FFASRHaebUmbach,XMM+15,Purushothaman2020}. Far-field speech signals such as for automatic meeting transcription and digital assistants normally require preprocessing to remove the detrimental effects of interference in the signal \cite{Haeb-Umbach2019, Hain3,close2022}. A number of methods have been proposed for speech dereverberation for both single channel and multichannel models \cite{Habets2010}. Recent advances in speech dereverberation performance in a number of domains have been driven by \ac{DNN} models \cite{dnnwpe,8553141,Wang2020DeepLB,Wang2021TeCANetTA,Zhao2021UNetBasedMS}.

%% Dereverberation Overview & Related Work
Convolutional neural network models are commonly used for sequence modelling in speech dereverberation tasks \cite{Wang2021ConvolutivePF,Fu2021DESNetAM,su20b_interspeech}.
One such fully convolutional model known as the \ac{TCN} has been proposed for a number of speech enhancement tasks \cite{WHAMR, 9095210,8683634}. \Acp{TCN} are capable of monaural speech dereverberation as well as more complex tasks such as joint speech dereverberation and speech separation \cite{9095210}.
The best performing \ac{TCN} models for speech dereverberation tasks typically have a larger receptive field for data with higher reverberation times T60 and a smaller receptive field for data with small T60s \cite{rfield} which forms the motivation for this paper.

%% In this work...
In this work, a novel \ac{TCN} architecture is proposed which is able to focus on specific temporal context within its receptive field. This is achieved by using an additional depthwise convolution kernel in the depthwise-separable convolution with a small dilation factor. Inspired by work in dynamic convolutional networks, an attention network is used to selected how to weight each of the depthwise kernels \cite{dynadnn,dynaconv}. 

%% The remainder of this paper...
The remainder of this paper proceeds as follows. Section~\ref{sec:WD-TCN} introduces the signal model and the \ac{WD-TCN} dereverberation network. Section~\ref{sec:exps} describes the experimental setup and data and results are presented in Section~\ref{sec:results}. Section~\ref{sec:conclusions} concludes the paper.

%%%%%%%%%%%%%%%%% SECTION %%%%%%%%%%%%%%%%% 
\section{Dereverberation Network}
\label{sec:WD-TCN}
In this section the monaural speech dereverberation signal model is introduced and the proposed \ac{WD-TCN} dereverberation model is described. The general \ac{WD-TCN} model architecture is similar to the reformulation of the Conv-TasNet speech separation model \cite{convtasnet} as a \ac{DAE} in \cite{rfield}.
%
%%% Signal Model %%%
\subsection{Signal Model}
A reverberant single-channel speech signal
% of $L_x$ samples at sample $i$ 
is defined as
\begin{equation} \label{eq:SigModel}
    x[i] = h[i]\ast s[i] = s_\mathrm{dir}[i] + {s_\mathrm{rev}[i]}
\end{equation}
for discrete time index $i$ where $\ast$ denotes the convolution operator, $h[i]$ denotes a \ac{RIR} and $s[i]$ denotes the clean speech signal. 
%In the data used in Section~\ref{sec:exps} of 
In this paper the target speech is $s_\mathrm{dir}[i]=\alpha s[i-\tau]$, i.e.~the clean signal convolved with the direct path of the 
%speech 
\ac{RIR} from speaker to receiver, expressed by the delay of signal travel from speaker to receiver $\tau$ and attenuation factor $\alpha$.

The mixture signal $x[i]$ is processed in $L_\vek{x}$ blocks 
\begin{equation}\label{eq:InputSignalBlock}
    \vek{x}_\ell = 
\left[x[0.5(\ell-1)L_{\mathrm{BL}}],\ldots, x[0.5(1+\ell) L_{\mathrm{BL}}-1]\right]
%.
\end{equation}
% where $\vek{x}_\ell$ is the $\ell${th} block
of $L_\mathrm{BL}$ samples with a 50\% overlap for frame index $\ell \in \{1,\ldots,L_\vek{x}\}$.

%%% Encoder %%%
\subsection{Encoder}
The encoder is a 1D convolutional layer with trainable weights $\mat{B}\in\Real^{L_\mathrm{BL}\times N}$, where $L_\mathrm{BL}$ and $N$ are the kernel size and number of output channels respectively. This layer transforms $\vek{x}_\ell$ into a set of filterbank features $\mat{w}_\ell$ such that
\begin{equation}\label{eq:encoder}
 \mat{w}_\ell=\mathcal{H}_\mathrm{enc}\left(\vek{x}_\ell\mat{B}\right){,}
\end{equation}
where $\mathcal{H}_\mathrm{enc}:\Real^{1\times N} \rightarrow \Real^{ 1\times N}$ is a ReLU activation function.
%

%%% Mask Estimation %%%
\subsection{Mask Estimation using \texorpdfstring{\Acp{WD-TCN}}{WD-TCNs} }
A mask estimation network is trained to estimate a sequence of masks $\mat{m}_\ell$ that filter the encoded features $\vek{w}_\ell$ to produce an encoded dereverberated signal defined as
\begin{equation}
    \mat{v}_{\ell}=\mat{m}_{\ell}\odot\mat{w}_\ell.
\end{equation}
% In previous work the authors defined a \ac{TCN} model for speech dereverberation \cite{rfield} derived from reformulating the Conv-TasNet speech separation model \cite{convtasnet} as a \ac{DAE}. 
The $\odot$ operator denotes the Hadamard product. A more detailed description of the \ac{TCN} used as a baseline in this paper 
% and how it deviates from \cite{convtasnet} 
is provided in~\cite{atttasnet}. Streaming implementations of these models are feasible but for this paper we focus on utterance-level implementations for brevity \cite{convtasnet}. 
%Further discussion of expanding to streaming implementations is provided in Section~\ref{sec:conclusions}.

The conventional \ac{TCN} consists of an initial stage which normalizes the encoded features $\vek{w}_\ell$ and reduces the number of features from $N$ to $B$ for each block using a \ac{P-Conv} bottleneck layer \cite{convtasnet}.
The \ac{TCN} is composed of a stack of $X$ dilated convolutional blocks that is repeated $R$ times. This structure allows for increasingly larger models with increasingly larger receptive fields \cite{rfield}. The \ac{D-Conv} layer in the blocks has an increasing dilation factor to the power of two for the $X$ blocks in a stack, i.e.~the dilation factors $f$ for each block are taken from the set $\{2^0,2^1,\ldots,2^{X-1}\}$ in increasing order. Fig.~\ref{fig:convblocks}~(a) depicts the convolutional block as implemented in \cite{speechbrain,atttasnet}. 
The convolutional block consist primarily of \ac{P-Conv} and \ac{D-Conv} layers with \ac{PReLU} activation functions \cite{prelu} and \ac{gLN} layers~\cite{convtasnet}. The \ac{P-Conv} and \ac{D-Conv} layers are structured to allow increasingly larger models to have larger receptive fields. Combining these two operations is an operation known as \ac{DS-Conv}~\cite{convtasnet}. More detailed definitions of \ac{P-Conv}, \ac{D-Conv} and \ac{DS-Conv} layers are given in Section~\ref{sec:dsconv} before the proposed \ac{WD-TCN} to replace the \ac{DS-Conv} operations in \acp{TCN} is introduced in Section~\ref{sec:multi_dilation_dsconv}, denoted in this paper as \ac{WD-Conv}.
%
%%% P-Conv vs D-Conv and DS-conv%%%
\subsubsection{Depthwise-Separable Convolution (DS-Conv)}
\label{sec:dsconv}
The \ac{DS-Conv} operation is a factorised version of standard convolutional kernel using a \ac{D-Conv} layer and a \ac{P-Conv} layer. The main motivation for using \ac{DS-Conv} is primarily parameter efficiency where the number of channels is sufficiently larger than the kernel size \cite{convtasnet}. Note that in this section the focus is entirely on 1D convolutional kernels but the same principle can be extended to higher dimensional kernels.

The \ac{D-Conv} layer is an entirely sequential convolution with dilation factor $f$, i.e.~each operation operates on each input channel individually. For the matrix of input features $\vek{Y}\in\Real^{G \times L_{\vek{x}}}$ where $G$ is the number of input channels (and consequently also the number of output channels) the \ac{D-Conv} operation can be defined as
\begin{equation}
    \mathcal{D}(\vek{Y},\vek{K}_{\mathcal{D}})=\left[(\vek{y}_0 * {\vek{k}}_0)^{\top},\ldots,({\vek{y}}_{G-1} * {\vek{k}}_{G-1})^{\top}\right]^{\top}
\end{equation}
where $\vek{K}_{\mathcal{D}}\in\mathbb{R}^{G \times P}$ is the the \ac{D-Conv} kernel matrix of trainable weights and $P$ is the kernel size. The $g$th row of $\vek{Y}$ and $\vek{K}_\mathcal{D}$ are denoted by $\vek{y}_g$ and $\vek{k}_g$ respectively. 

The \ac{P-Conv} layer is an entirely channel-wise convolution. This operation in practice is a standard 1D convolutional kernel but with only a kernel size of 1. The \ac{P-Conv} operation can be defined as
\begin{equation}
    \mathcal{P}(\vek{Y},\mathbf{K}_{\mathcal{P}})=\vek{Y}^\top%\ast
    \vek{K}_{\mathcal{P}}
\end{equation}
where $\mathbf{K}_{\mathcal{P}}\in\mathbb{R}^{G\times H}$ is the \ac{P-Conv} kernel of trainable weights.

\begin{figure}[!ht]
    \centering
    \includegraphics[width=\columnwidth]{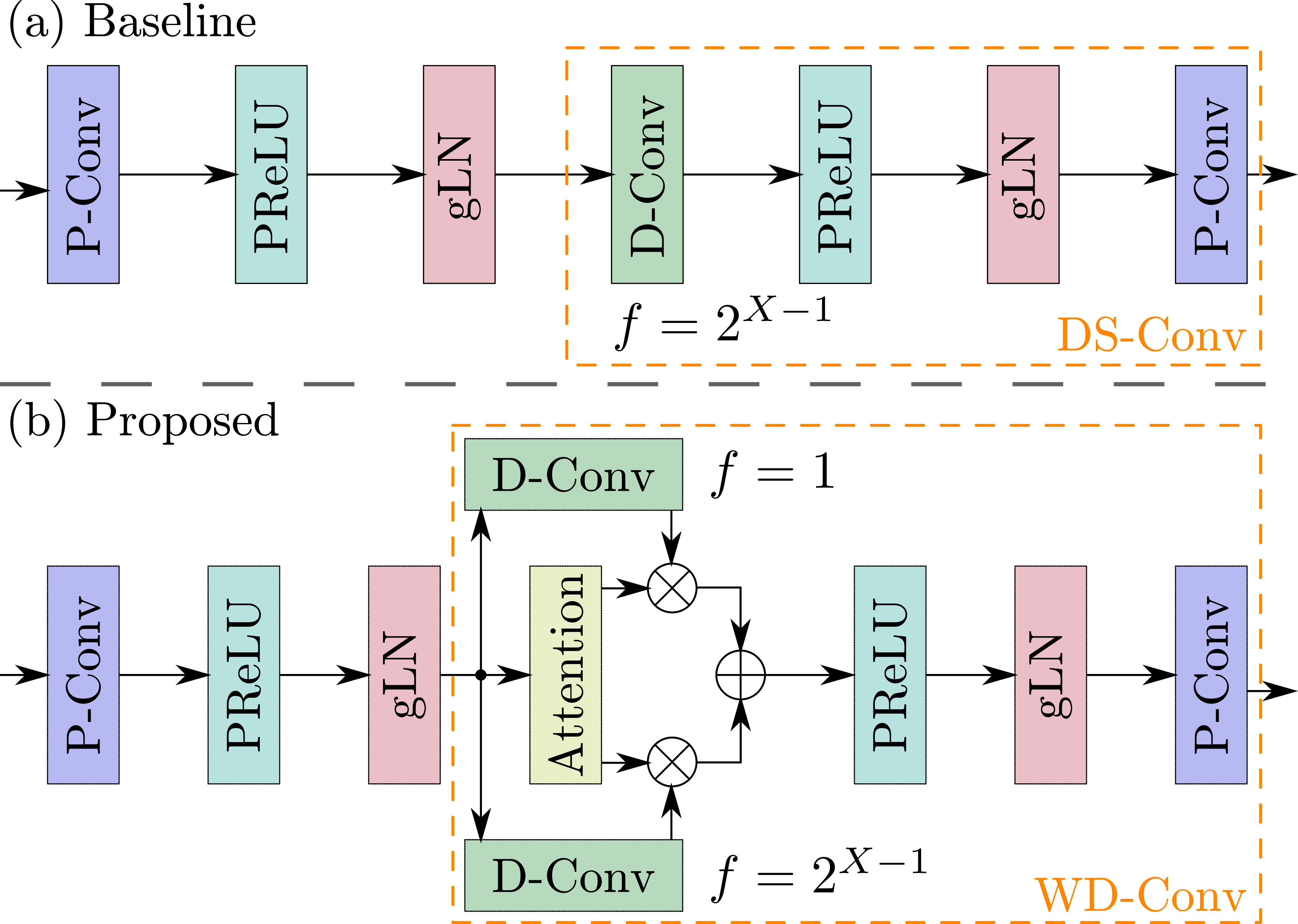} 
    \caption{(a) Convolutional block in baseline \ac{TCN}; (b) Proposed convolutional block. Example for final block in a stack of conv.~blocks for $Q=2$ with dilation factor $f=2^{X-1}$. Note that a residual connection around the entire block is omitted for brevity.}
    \label{fig:convblocks}
\end{figure}

Combining the definitions for the \ac{D-Conv} and \ac{P-Conv} operations, the \ac{DS-Conv} operation is defined as
\begin{equation}
    \mathcal{S}\left(\vek{Y},\vek{K}_{\mathcal{D}},\mathbf{K}_{\mathcal{P}}\right)=\mathcal{P}\left(\mathcal{D}\left(\vek{Y},\vek{K}_{\mathcal{D}}\right),\mathbf{K}_{\mathcal{P}}\right).
\end{equation}
The \ac{DS-Conv} operation as implemented in the baseline system used in \cite{rfield} and in this paper can be seen in Fig.~\ref{fig:convblocks}~(a) highlighted by the dashed orange box.

\subsubsection{Weighted Multi-Dilation Depthwise-Separable Convolution (WD-Conv)}
\label{sec:multi_dilation_dsconv}
The \ac{WD-Conv} network structure depicted in Fig.~\ref{fig:convblocks}~(b) is proposed here as an extension to the \ac{DS-Conv} operation where it is preferable to allow the network to be more selective about the temporal context to focus on without drastically increasing the number of parameters in the model. The proposed \ac{WD-Conv} layer incorporates additional parallel \ac{D-Conv} layers that can have a different dilation factor, hence it is referred to as dilation-augmented. The output of the \ac{D-Conv} layers are weighted in a sum-to-one fashion and summed together. This summed output is then passed as the input to a \ac{P-Conv}. In its simplest form the \ac{WD-Conv} operation can be formulated as
\begin{multline}
    \label{eq:wdconf}
    \mathcal{W}\left(\vek{Y},\left(\vek{K}_{\mathcal{D}_1,},\ldots,\vek{K}_{\mathcal{D}_Q}\right),\mathbf{K}_{\mathcal{P}}\right)=\\
    \mathcal{P}\left(\sum_{q=1}^Q a_q\mathcal{D}_q\left(\vek{Y},\vek{K}_{\mathcal{D}_q}\right),\mathbf{K}_{\mathcal{P}}\right)
\end{multline}
where $Q$ is the number of parallel \acp{D-Conv} in the \ac{WD-Conv} and $a_q$ are their corresponding weights that sum-to-one, i.e.~$\sum_{q=1}^Q a_q=1$. In the model proposed here the number of \acp{D-Conv} is set to $Q=2$ ; one with a dilation factor $f=1$ and the other according to the exponentially increasing dilation rule defined previously and used in \cite{convtasnet,rfield,atttasnet} where $f$ is increasing in powers of 2 with every successive block in a stack of $X$ blocks. Note that the first convolutional blocks of a stack of $X$ blocks in the proposed implementation use an identical dilation of $f=1$ for each of the \ac{D-Conv} kernels in the \ac{WD-Conv} operation.

\begin{figure}[!t]
    \centering\large
    \includegraphics[width=\columnwidth]{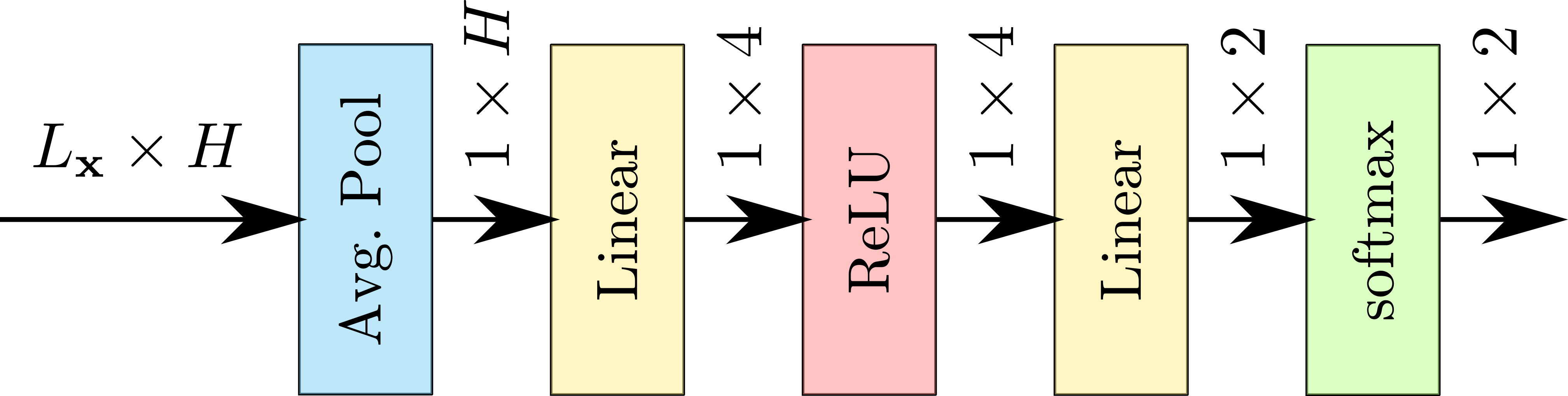} 
    \caption{Squeeze and excite attention weighting network. Output dimensionality of each layer is indicated above arrows.}
    \label{fig:se_network}
\end{figure}

Inspired by the dynamic convolution kernel proposed in \cite{dynaconv}, the implementation proposed in this paper computes the weights for each \ac{D-Conv} layer using an \ac{SE} attention network\cite{8578843}. The \ac{SE} attention network  is shown in Fig.~\ref{fig:se_network} and is composed of a global average pooling layer that reduces the sequence dimension from $L_\vek{x}$ to $1$ producing a vector of dimension $H$, the same as the feature dimension of the input. This feature vector is then compressed using a linear layer, with a \ac{ReLU} activation, to a dimension of $4$ as in \cite{dynaconv}. The final stage is a linear layer that  computes a weight for each of the \ac{D-Conv} kernels in the \ac{WD-Conv} structure. In the proposed model there are two \ac{D-Conv} kernels and so the linear layer has an input dimension of $4$ and an output dimension of $2$. A softmax activation is used to ensure the sum-to-one constraint on the weights of the \ac{D-Conv} layers.
\subsection{Decoder}
The decoder tansforms the encoded dereverberated signal $\vek{v}_{\ell}$ back into a time domain signal using a transposed 1D convolutional layer with $N$ input channels, $1$ output channel and a kernel size of $L_\mathrm{BL}$ such that
\begin{equation}
 \hat{\vek{s}}_{\ell}=%(\mat{m}_{\ell}\odot\mat{w}_\ell)\mat{U}=
 \vek{v}_{\ell}\vek{U}
\end{equation}
where $\vek{U}\in\Real^{N\times L_\mathrm{BL}}$ is a matrix of trainable convolutional weights and $\hat{\vek{s}}_{\ell}$ is an estimated dereverberated signal block in the time domain. The overlap-add method is used for re-synthesis of the signal from the overlapping blocks.
\subsection{Objective Function}
The objective function used here is the \ac{SISDR} function \cite{LeRoux} which is the same as that used to train the baseline \ac{TCN} \cite{rfield}. It is reformulated as a loss function by taking the negative \ac{SISDR} value between the estimated speech segment $\hat{\vek{s}}$ and the reference direct path of the signal $\vek{s}_\mathrm{dir}$ defined as
\begin{equation} 
\label{eq:DefSISDR}
\mathcal{L}_\text{SISDR}(\hat{\vek{s}},\vek{s}_\mathrm{dir}): 
= - 10\log_{10} \frac{\left\Vert 
\frac{\langle \hat{\vek{s}},\vek{s}_\mathrm{dir}\rangle 
\vek{s}_\mathrm{dir}}{\Vert \vek{s}_\mathrm{dir}\Vert^{2}}\right\Vert^{2}}{\left\Vert\hat{\vek{s}}-\frac{\langle \hat{\vek{s}},\vek{s}_\mathrm{dir}\rangle 
\vek{s}_\mathrm{dir}}{\Vert \vek{s}_\mathrm{dir}\Vert^{2}}\right\Vert^{2}}.
\end{equation}

%%%%%%%%%%%%%%%%% SECTION %%%%%%%%%%%%%%%%% 
\section{Experimental Setup}
\label{sec:exps}

\subsection{Model and Training Configuration}
Different model configurations are compared in the following to demonstrate the improvement gained by the proposed \ac{WD-TCN} model across a range of model sizes. This is done by varying the number of convolutional blocks in a dilated stack $X$ as well as the number of times the dilated stack is repeated $R$. Based on previous work \cite{rfield}, the ranges of $X\in\{4,5,6,7,8\}$ and $R\in\{4,5,6,7,8\}$ were selected, resulting in 25 different configurations. All other parameters are fixed, i.e.~kernel size $L_\mathrm{BL}=16$,  number of encoder output channels $N=512$, number of bottleneck output channels $B=128$, number of channels inside the convolutional blocks $H=512$ and the kernel size inside each \ac{D-Conv} $P=3$. For more details on these parameters see \cite{rfield,atttasnet,convtasnet}.

The same training approach as in \cite{rfield} is used for both the baseline \ac{TCN} and \ac{WD-TCN}. Each model is trained for 100 epochs. An initial learning rate of $0.001$ is used and is halved if there is no improvement for 3 epochs. A batch size of 4 is used.
The training was performed using the SpeechBrain speech processing toolkit \cite{speechbrain}. The implementation of the proposed \ac{WD-TCN} is available on GitHub\footnote{Link to \ac{WD-TCN} model on GitHub: \url{https://github.com/jwr1995/WD-TCN}}.
\subsection{Data}
The simulated WHAMR noisy reverberant two speaker speech separation corpus \cite{WHAMR} is used for the following experiments in this section. Only the reverberant and clean first speaker data is used for the input data $x[i]$ and target data $s_\mathrm{dir}[i]$. \Acp{RIR} are simulated using the pyroomacoustics software toolkit \cite{pyroomacoustics} and then convolved with the speech clips to produce the reverberant signal $x[i]$. The training set contains 20,000 samples for training which are truncated or padded to 4~s in length, to address sample length mismatches in batches and to also speed up training. There are 5000 samples (14.65~hrs) and 3000 samples (9~hrs) in the validation and test sets respectively.

\subsection{Metrics}
%% Metrics
A number of metrics are used to assess a variety of properties in the dereverberated speech.
The objective function \ac{SISDR} is also used to measure distortions in signals.
\Ac{SRMR}  \cite{srmr} is a measure used to directly measure reverberant effects in the signal. 
\Ac{PESQ} \cite{PESQ} and \ac{ESTOI} \cite{estoi} are objective measures used to assess the quality and intelligibility of signals. 
%In addition, $\Delta$ and improvement measures are used. Measures denoted with $\Delta$ indicate change in measurement value between the reverberant signal $x[i]$ and the dereverberated signal $\hat{s}_\mathrm{dir}[i]$. Improvement measures indicate the improvement of the \ac{WD-TCN} model over the \ac{TCN} baseline from \cite{rfield}. 
%\textcolor{orange}{ESTOI NOT USED. Rewirte $\Delta$ and improvement part.}
%
\section{Results}
\label{sec:results}
\subsection{Performance Metrics and Model Size}
The average \ac{SISDR} results on the WHAMR evaluation set for the 25 chosen model configurations of the proposed \ac{WD-TCN} model are given in \autoref{tab:tcn_se} with \ac{SISDR} improvements over the \ac{TCN} model in the parenthesis. The bold font 
%in Table~\ref{tab:tcn_se} 
indicates best performance and highest improvement, respectively. 
\setlength{\tabcolsep}{3pt}
\begin{table}[!b]
\centering
\resizebox{\columnwidth}{!}{%
\begin{tabular}{cc|ccccc}
 &  & \cellcolor[HTML]{C0C0C0} & \cellcolor[HTML]{C0C0C0} & \cellcolor[HTML]{C0C0C0}$X$ & \cellcolor[HTML]{C0C0C0} & \cellcolor[HTML]{C0C0C0} \\ \cline{3-7} 
 &  & \cellcolor[HTML]{C0C0C0}4 & \cellcolor[HTML]{C0C0C0}5 & \cellcolor[HTML]{C0C0C0}6 & \cellcolor[HTML]{C0C0C0}7 & \cellcolor[HTML]{C0C0C0}8 \\ \hline
\multicolumn{1}{l|}{\cellcolor[HTML]{C0C0C0}} & \cellcolor[HTML]{C0C0C0}4 & 11.21 (.28) & 11.66 (.29) & 11.81 (.40) & 11.94 (.38) & 12.04 (.40) \\
\multicolumn{1}{l|}{\cellcolor[HTML]{C0C0C0}} & \cellcolor[HTML]{C0C0C0}5 & 11.51 (.41) & 11.86 (.41) & 11.94 (.23) & 12.11 (.42) & 12.11 (.39) \\
\multicolumn{1}{l|}{\cellcolor[HTML]{C0C0C0}$R$} & \cellcolor[HTML]{C0C0C0}6 & 11.64 (.38) & 11.95 (.30) & 12.08 (.31) & 12.09 (.23) & 12.11 (.20) \\
\multicolumn{1}{l|}{\cellcolor[HTML]{C0C0C0}} & \cellcolor[HTML]{C0C0C0}7 & 11.65 (.20) & 12.17 (\textbf{.44}) & 12.22 (.30) & 12.16 (.13) & 12.14 (.16) \\
\multicolumn{1}{l|}{\cellcolor[HTML]{C0C0C0}} & \cellcolor[HTML]{C0C0C0}8 & 11.79 (.27) & 12.03 (.19) & 12.20 (.17) & 12.21 (.22) & \textbf{12.26} (.32)
\end{tabular}
}
\caption{\ac{SISDR} performance of \Ac{WD-TCN} with \ac{SE} attention in dB. Numbers in $(\cdot)$ report performance improvement over baseline TCN.}
\label{tab:tcn_se}
\end{table}
These results show that the \ac{WD-TCN} outperforms the \ac{TCN} model across all 25 model configurations. The biggest performance gains are seen around $\{X,R\} = \{5,7\}$ and the \ac{WD-TCN} model with most parameters and largest receptive field, $\{X,R\}=\{8,8\}$, shows best overall performance, contrary to the \ac{TCN} model which gave the best SISDR results with $\{X,R\}=\{6,8\}$. 
\begin{figure}[!t]
    \centering
    \includegraphics{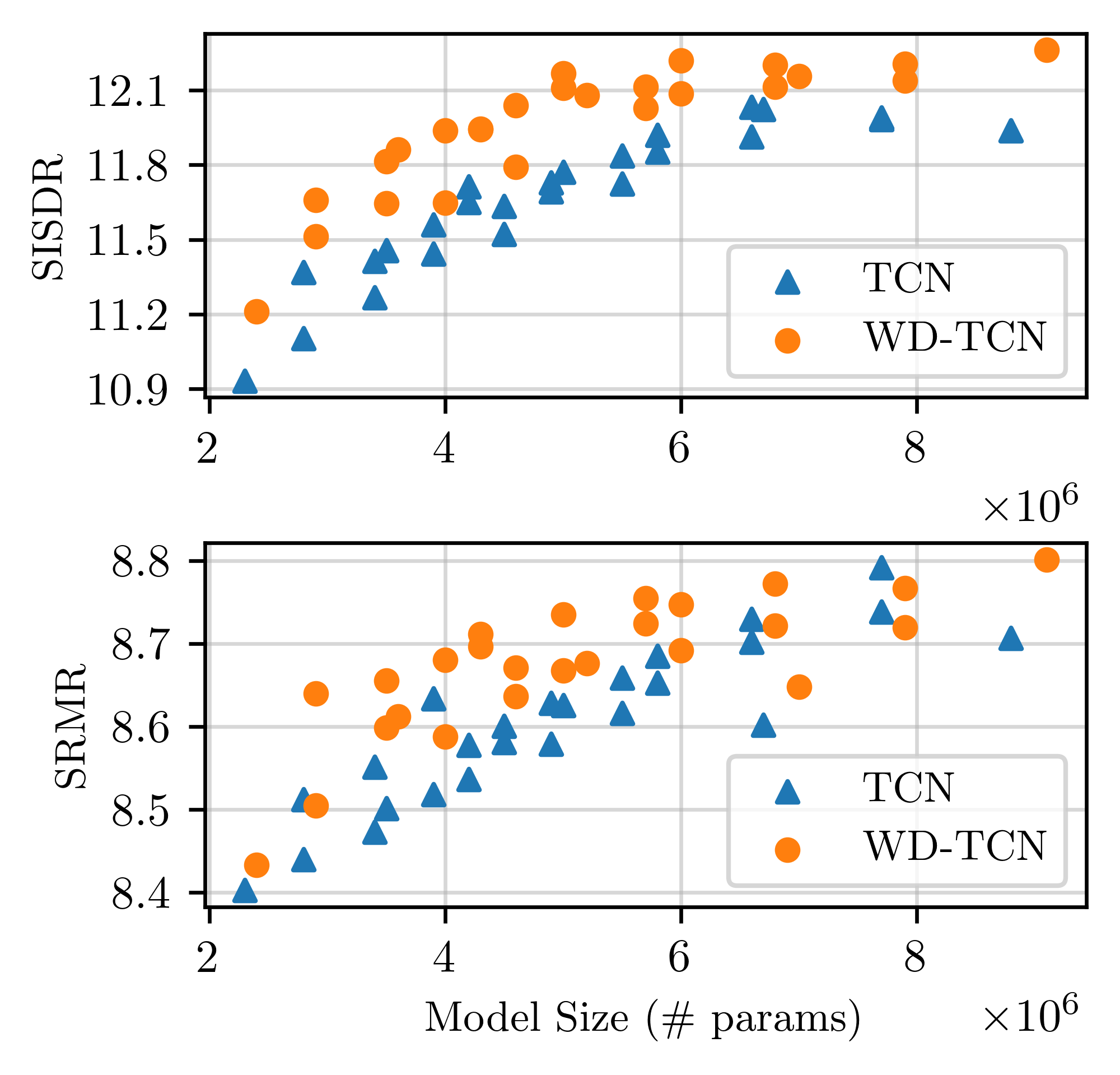}
    \caption{Comparison of baseline \ac{TCN} and \ac{WD-TCN} over model size (no.~of parameters) in terms of 
    %$\Delta$~
    \ac{SISDR} (top) and 
    %$\Delta$~
    \ac{SRMR} (bottom).}
    \label{fig:nparams}
\end{figure}
Fig.~\ref{fig:nparams}~(top) shows 
%$\Delta$~
\ac{SISDR} performance for all models over the model sizes in number of parameters. It can be seen that using the \ac{WD-TCN} is a more parameter efficient approach to improving model performance than increasing the number of convolutional blocks (larger $X$ or $R$ values) in a conventional \ac{TCN}.
The 
%$\Delta$ 
\ac{SRMR} performance against model size (Fig.~\ref{fig:nparams}, lower panel) shows the same findings, i.e.~that the \ac{WD-TCN} is a more parameter efficient approach to improving performance. For some larger models ($>$~6M parameters) performance differs less. However the best performing model in terms of %$\Delta$ 
\ac{SRMR} is still the \ac{WD-TCN}.

\autoref{tab:full_measures} shows the results of the best performing TCN and WD-TCN models for each of the chosen performance metrics, highlighted in yellow, compared with the respective other model for the same $X$ and $R$ hyper-parameters. 
% Table \ref{tab:full_measures} shows the results for the best performing TCN and WD-TCN models compared with their corresponding models in terms of the configuration of convolutional blocks determined by $X$ and $R$. 
%
The performance in PESQ is inconclusive as many \ac{TCN} models outperform their corresponding \ac{WD-TCN} configurations but the best PESQ score of 3.5 is achieved with the \ac{WD-TCN} model. The \ac{WD-TCN} models show slightly better performance in ESTOI in line with the trend already observed in SRMR and SISDR. Note that SRMR is considered the most significant metric as it is designed to assess reverberation only.
%The performance increase in \ac{SRMR} and \ac{SISDR} is also mostly reflected in \ac{ESTOI} but these improvements are very minor. Many \ac{TCN} models outperformed the corresponding \ac{WD-TCN} models in terms of \ac{PESQ} however the best overall performing model in \ac{PESQ} was a \ac{WD-TCN} model with configuration $\{X,R\}=\{6,7\}$.
\input{full_measures}
\subsection{Squeeze-and-Excite Attention Analysis and T60 Variation}
\begin{figure}[!ht]
    \centering
    \includegraphics[width=\columnwidth]{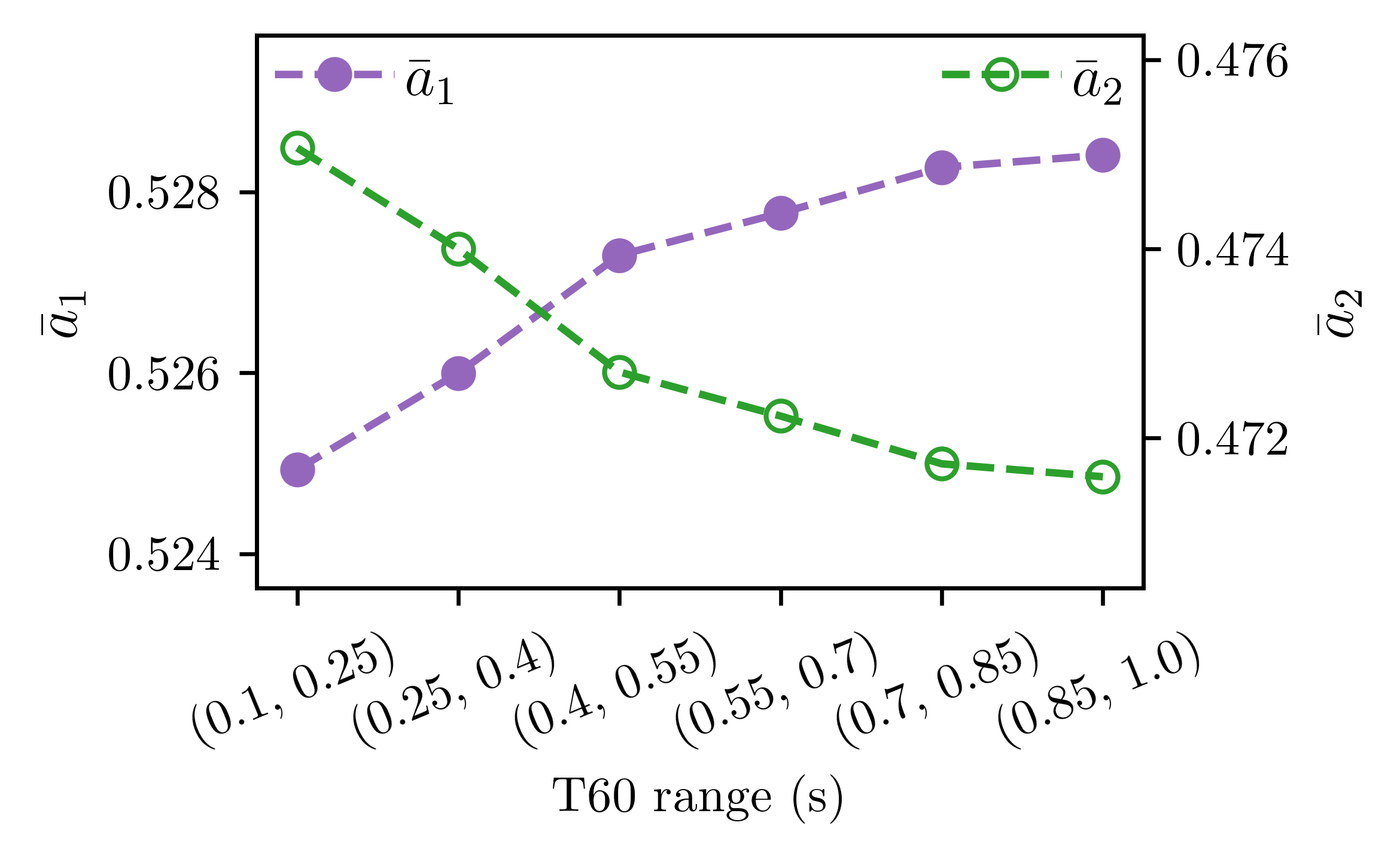}
    \caption{Mean values of attention weights $\bar{a}_q$ across six different T60 ranges in the WHAMR evaluation set over all models with $X\in\{4,\ldots,8\}$ and $R\in\{4,\ldots,8\}$.}
    \label{fig:meanvsT60}
\end{figure}

In the following, the attentive weights $a_q$ in (\ref{eq:wdconf}) in the convolutional blocks are analysed. Note that $a_1$ corresponds to the attention weight applied to the \ac{D-Conv} layers with the increasing dilation of $f\in\{1,2,\ldots,2^{X-1}\}$ for all convolutional blocks (cf.~Fig.~\ref{fig:convblocks}~(b)) and $a_2=$ is the weight corresponding to the \ac{D-Conv} layers with the more local fixed dilation $f=1$.
%
% The variation of $a_1$ for three increasingly larger models size and for 3 speech samples of increasing T60 is shown in Fig.~\ref{fig:xranalysis}. Many blocks were trained such that the attention gave balanced mixtures of the two \acp{D-Conv}, i.e. $a_1 \approx a_2 \approx 0.5$. It is also noticeable that for the larger the model, the greater the variance of $a_1$ appears to be. It was also observed that samples with low T60 values typically produce less variance in $a_1$ as the overall model size increases.
% %
% \begin{figure}[!ht]
%     \centering
%     \includegraphics{X_R_analysis.png}
%     \caption{Variation of $a_1$ through the \ac{WD-TCN} for three different samples of varying T60 values for three different model configurations $\{X,R\}=\{4,4\}$ (top), $\{X,R\}=\{6,6\}$ (middle) and $\{X,R\}=\{8,8\}$ (bottom). Vertical dashed lines denote blocks with the highest dilation factors $f=2^{X-1}$.}
%     \label{fig:xranalysis}
% \end{figure}
%
% This is demonstrated in Fig.~\ref{fig:stdvsT60} where it is shown that for larger ranges of T60 in the WHAMR evaluation set the mean standard deviation of $a_1$, denoted as $\bar{\sigma}_{a_1}$, is larger.
% %
% \begin{figure}[!ht]
%     \centering
%     \includegraphics{std_T60_5d.png}
%     \caption{Mean standard deviation of $a_1$ (denoted $\bar{\sigma}_{a_1}$) across five different T60 ranges in the WHAMR evaluation set for 5 models of different size (indicated by their $\{X,R\}$ parameter pair).}
%     \label{fig:stdvsT60}
% \end{figure}
%
To analyse whether the \ac{SE} attention approach was working as intended the attention weights were firstly computed across the entire evaluation set for every WD-TCN model trained in  \autoref{tab:tcn_se}. Mean values of the weights for each model and each sample in the evaluation set were then computed and the evaluation set was in divided into increasing T60 ranges from 0.1s up to 1s. The mean for each  weight $a_q$ over all models and samples, denoted as $\bar{a}_q, q\in \{1,2\}$, was then computed for each T60 range. Figure \ref{fig:meanvsT60} shows how the mean weight values vary across increasing T60 ranges. As the T60 range increases $\bar{a}_1$ increases. This demonstrates the \ac{SE} attention approach is working as intended because the network has a less local focus within its receptive field for speech signals with larger reverberation times. Similarly the mean of the local attention weight $\bar{a}_2$ decreases as the T60 range increases demonstrating that the network is more focused on local information in its receptive field when the speech has a smaller reverberation time.
% \begin{figure}[!ht]
%     \centering
%     \includegraphics[width=\columnwidth]{mean_T60_6d.png}
%     \caption{Mean values of attention weights $a_q$ (denoted $\bar{a_q}$) across six different T60 ranges in the WHAMR evaluation set over all models with $X\in\{4,\ldots,8\}$ and $R\in\{4,\ldots,8\}$.}
%     \label{fig:meanvsT60}
% \end{figure}
%
%%%%%%%%%%%%%%%%% SECTION %%%%%%%%%%%%%%%%% 
\section{Conclusions}
\label{sec:conclusions}
%% General Conclusions
In this work, the \ac{WD-TCN} model was proposed for \ac{TCN}-based speech dereverberation by replacing depthwise-separable convolutions with weight multi-dilation depthwise-separable convolutions. It was shown that the \ac{WD-TCN} consistently outperformed a conventional \ac{TCN} across 25 different model configurations and that using the \ac{WD-TCN} was a more parameter efficient approach to improving model performance than increasing the number of convolutional blocks in the \ac{TCN}.
\bibliographystyle{IEEEbib}
\bibliography{refs}

\end{document}

%% file: full_measures.tex
\begin{table}[!ht]
\begin{tabular}{|c|c|c|c|cccc|}
\hline
\rowcolor[HTML]{C0C0C0} 
\textbf{Model} & \textbf{X} & \textbf{R} & \textbf{\# params} & \textbf{SISDR} & \textbf{PESQ} & \textbf{ESTOI} & \textbf{SRMR} \\ \hline
TCN & 6 & 7 & 5.8M & 11.92 & 3.46 & 0.930 & 8.65 \\
WD-TCN & 6 & 7 & 6.0M & \textbf{12.22} & \cellcolor[HTML]{FFFFC7}\textbf{3.5} & \textbf{0.933} & \textbf{8.69} \\ \hline
TCN & 6 & 8 & 6.6M & \cellcolor[HTML]{FFFFC7}12.03 & \textbf{3.46} & 0.932 & 8.70 \\
WD-TCN & 6 & 8 & 6.8M & \textbf{12.20} & 3.43 & \textbf{0.934} & \textbf{8.72} \\ \hline
TCN & 8 & 4 & 4.5M & 11.63 & \cellcolor[HTML]{FFFFC7}\textbf{3.48} & 0.927 & 8.60 \\
WD-TCN & 8 & 4 & 4.6M & \textbf{12.04} & 3.45 & \textbf{0.931} & \textbf{8.67} \\ \hline
TCN & 8 & 7 & 7.7M & 11.98 & \textbf{3.46} & 0.933 & \cellcolor[HTML]{FFFFC7}\textbf{8.79} \\
WD-TCN & 8 & 7 & 7.9M & \textbf{12.14} & 3.45 & \textbf{0.935} & 8.72 \\ \hline
TCN & 8 & 8 & 8.8M & 11.94 & \textbf{3.46} & \cellcolor[HTML]{FFFFC7}0.933 & 8.71 \\
WD-TCN & 8 & 8 & 9.1M & \cellcolor[HTML]{FFFFC7}\textbf{12.26} & 3.45 & \cellcolor[HTML]{FFFFC7}\textbf{0.935} & \cellcolor[HTML]{FFFFC7}\textbf{8.8} \\ \hline
\end{tabular}
\caption{Best performing TCN and WD-TCN models compared corresponding models in SISDR, PESQ, ESTOI and SRMR. Bold indicates best performance per configuration, in terms of the $X$ and $R$ hyper-parameters. Results highlighted in yellow indicate best overall results for each model in each metric.}
\label{tab:full_measures}
\end{table}